# Realized collaboration across borders - Linux and the bioinformatic part

Sasa Paporovic

Email: sasa.paporovic@fh-bielefeld.de

May 19, 2013

v1.0

In year 2006 Biolinux inter alia with the work of Tim Booth gives its rise and provide an operating system that was and is specialized in providing an environment for the needs of bioinformatics[1].

Since than some years have been seen and the information about Biolinux has not only widespread within the bio-science community, also the Linux community at whole had a ear for it.

This has caused a process that is unique in the Linux community where separation from each other is a profile issue of the different Linux projects[2,3].

The process is called "collaboration across borders" and makes as the keywords say the different work groups cooperating together[4].

In many fields like the different Linux desktops(KDE,GNOME,... other) this is failing[2,3,5].
In bioinformatics it is working.

In this paper we will have a look into the history of this collaboration, will show its growth until we reach the actual state and we will talk about the "quo vadis" of this process.

**Where the Linux come from**

Biolinux is not a project thats starts de novo in gathering software and construct a Linux. In fact Biolinux has its successors, not in a sense of full substitution, more in a sense of a growing family.

In 1991 Linux was designed and in its first version(only) programmed by Linus Torvalds[6,7].

As can be seen in the original message by Torvalds[7] he himself saw his own doing as a little project and the text gives no hint, that he expected in any way the proliferation of his work to the state we have today, in which Linux is to everybody a good known term, even when Linux has only a 1%-2% worldwide market share on Desktop PC's[8,9]. But on server systems it is estimated between 30%-70%[10,11], which makes it important to everyone, especially in the sense that the Internet is build of servers as usual as Linux.

Anyway, in the proliferation and growing process Linux had some remarkable points. On of them we have already mentioned. The dawn of Biolinux in 2006.

The other which were important in the view towards the development of Biolinux we are going to facing now.

**In 1993 the Debian project had its birth[12].**

The two years after the founding of the Linux project, Linux was only usable to very keen experts. One of the causes for this was that Linux had no compilation, or better distribution process, which suited the needs for a ready and stable system[13].

This was changed by the Debian project. Some developers has recognized this bottleneck and decided to cooperate with each other to bring Linux in a state which could fit the slogan "A Linux for everybody". They have had success with their approach and Debian in its first version was born.

Afterwards it happened what happened. Linux growth, Debian gets its next version and all geeks entered a new metaplane of happiness.

But not for all people this development state was already good enough.

**Ubuntu- the journey to the geek haven**

Mark Shuttleworth have felt a little bit uncomfortable with the technical progress that the Debian project have had. You know, all in Debian is of stability of the system and for getting this on in the stable versions of Debian it is not always the very latest software in there.

To getting around this Ubuntu was founded 2004[14].

It imports its software from the Debian project, but not from the stable branch. Instead it derives its software from the testing branch of Debian[15], which is the Debian version that will come next and is still under development[16].

That was keen.

Using the work that was already done by the people of Debian was absolutely in the sense of open source software, and also it was absolutely in the sense of open source software to give back.
This was and is done by the Ubuntu people. Debian gives the groundwork and Ubuntu gives some refinements back to get the next version of Debian to stable during their release preparation.

"And the geek haven?"

Is not reached at all, but we see the thinks going forward. Ubuntu is getting nearer and nearer to the latest software without remarkable loose of stability.

**Biolinux – some geeks are very special.**

In 2006 some people, a Tim Booth was one of them felt was geeks are feeling from time to time. Unsatisfactory with the technical possibilities.
So, the story went on. They do the same as the Ubuntu people do with Debian. They derive from Ubuntu, performed some special hacker key strokes and call the result Biolinux, a Linux that is specialized in providing an environment and a software compilation for bioinformatics[1].

And, sure, in the best sense of whiteheads[17] they give back from their work.

So the cycle closes

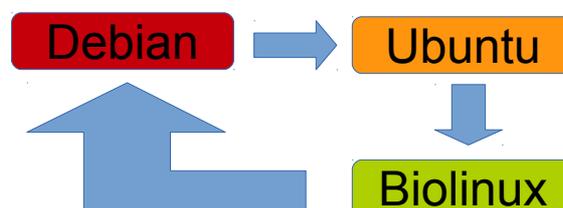

*Graphic 1: Genealogie of Debian derived Linux distributions and the work cycle for the bioinformatic software integration. Graphic designed with LibreOffice Draw 4.0.2*

**Software is like plants are – computer scientists also**

To shorten the story.
There was growth . Absolutely not in a ordered sense of a Japanese garden[18] or in the good old English one[19]. It was more in a sense of, well, look in the garden of your next best neighbor. With high probability you will get a clue what is meant.

Today you can meet Tim Booth from Biolinux at the mailing list of the Debian-Med project[20] right beside Anreas Tille, one of the founder of Debian Med[21]. The other way around on Debian-Med Alioth[22] are entries from and for the Biolinux project, as also an eye kept on Bugs that occur on launchpad[23], the collaboration platform for and around Ubuntu.

You can find the Debian Med team[24] as also the Biolinux Team[25] on launchpad, where they have their outpost for Ubuntu to get in touch with Ubuntu ecosystem and the people contributing from there.

It looks approximately like:

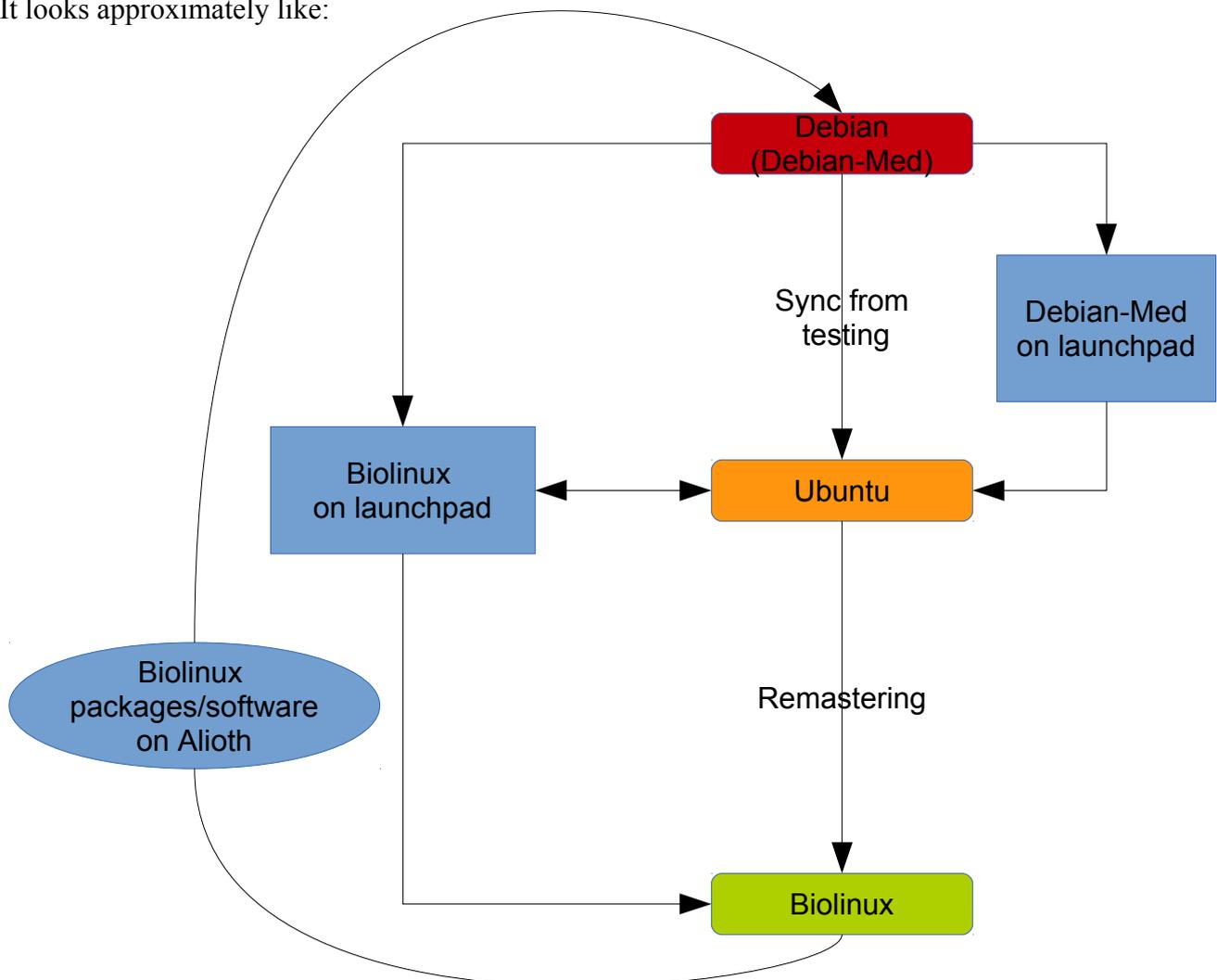

Graphic 2: Genealogy of Debian derived Linux distributions and the work cycle for the bioinformatic software integration, enriched with the involved work groups. Graphic designed with LibreOffice Draw 4.0.2

**How it comes to this extraordinary collaboration within the Linux world?**

Unfortunately the author has no good answer on this.

The causes may lay deep in the archives of the mailing list of the involved projects or in privat communications between the protagonists.

In any case it could be of no doubt that everything in past must have been reside of a foundation of respect to the each others work.

May be the collaborations has been forced by the huge problems that computer scientists in bioinformatics are facing.

May be, it was just no time there to spend in wounded pride and vanity. Good science was more important.

As Randy Pausch says in his "Last Lecture", sometimes it is about simple asking[26].

Following this a simple E-Mail to the protagonists is in preparation that will ask them how they see the causes for the collaboration observed.

**"Quo vadis?"**

Andreas Tille and team are providing a continiously stream of software selected by them to Debian. The Ubuntu project is discussing a shortening of the release cycles to bring the latest software much even faster to the users[27], and Tim Booth with team have extended their system integration work with giving some candies to the dafault software compilation[28](see Table1)

| Package | Version | Description |
| --- | --- | --- |
| backups | 0.3-6 | Bio-Linux 5/6 backup utilities. |
| base-directories | 1.0-37 | Essential Bio-Linux configuration files. |
| bldp-files | 1.1-96 | Documentation for bioinformatic software on Bio-Linux. |
| cloudbl-desktop | 1.4 | Adds usefull links to the remote desktop for Cloud BL |
| keyring | 5.0 | Public key package for the Bio-Linux package repository |
| plymouth-theme | 7.0 | Boot splash for Bio-Linux 7 |
| prevent-upgrade | 0.2 | Stop Bio-Linux from upgrading on release of 12.04 LTS |
| shared | 1.1-2 | Some shared files used by Bio-Linux |
| themes | 6.1 | Graphics, icons, wallpaper and a GDM theme for Bio-Linux 6 |
| themes -v5 | 1.2 | Graphics, icons, KDE splash screen and a GDM theme for Bio-Linux 5. |
| themes-v7 | 7.0-11 | Grahics, icons, wallpaper and desktop settings for Bbio-Linux 7 |
| tutorials | 1-23 | Tutorials and documentation for Bio-Linux users. |
| unity-lens | 0.2 | unity lens for Bio-Linux |
| usb-maker | 7.0-11 | Script to generate Bio-Linux live memory sticks. |

*Table 1: Biolinux core system specific packages. Reproduced from[28]. Table designed with LibreOffice Calc 4.0.2*

All is running and speeding up.

And may be, may be it is possible to reach other parts of the Bioinformatics Linux community(VLinux[29],openSUSE-Medical[30]) even if they use a complete different software managment system. Converters like alien are available[31].

Finis